Smartphone-based Optical Sectioning (SOS) Microscopy with A Telecentric Design for Fluorescence Imaging

Ziao Jiao[a,b,*], Mingliang Pan [a], Khadija Yousaf [c], Daniel Doveiko [c], Michelle Maclean[a,d], David Griffin[a], Yu Chen [c], and David Day Uei Li[a,b]

[a] Department of Biomedical Engineering, University of Strathclyde, Glasgow G1 1XQ, Scotland, UK
[b] Strathclyde Institute of Pharmacy and Biomedical Sciences, University of Strathclyde, Glasgow G4 0RE, Scotland, UK
[c] Department of Physics, University of Strathclyde, Glasgow G4 0RE, Scotland, UK
[d] Department of Electronic & Electrical Engineering, The Robertson Trust Laboratory for Electronic Sterilisation Technologies (ROLEST), University of Strathclyde, Glasgow, UK
[*] ziao.jiao@strath.ac.uk

**Abstract**
We proposed a Smartphone-based Optical Sectioning (SOS) microscope based on the HiLo technique, with a single smartphone replacing a high-cost illumination source and a camera sensor. We built our SOS with off-the-shelf optical mechanical cage systems with 3D-printed adapters to integrate the smartphone with the SOS main body seamlessly. The liquid light guide can be integrated with the adapter, guiding the smartphone's LED light to the digital mirror device (DMD) with neglectable loss. We used an electrically tunable lens (ETL) instead of a mechanical translation stage to realize low-cost axial scanning. The ETL was conjugated to the objective lens's back pupil plane (BPP) to construct a telecentric design by a 4f configuration. This can exempt images of different layers from the variation in magnification. SOS has a 571.5 μm telecentric scanning range and an 11.7 μm axial resolution. The broadband smartphone LED torch can effectively excite fluorescent polystyrene (PS) beads. We successfully used SOS for high contrast fluorescent PS beads imaging with different wavelengths and optical sectioning imaging of accumulated fluorescent PS beads. To our knowledge, the proposed SOS is the first smartphone-based HiLo optical sectioning microscopy. It is a powerful, low-cost tool for biomedical research in resource-limited areas.

1. Introduction

Widefield fluorescence microscopy (WFM) is commonly used to image biological samples due to its high cost-effectiveness, fast imaging speed, low photodamage, and low photobleaching [1–3]. However, according to WFM's optical transfer function (OTF), the optical sectioning ability is poor, which means WFM images have a low contrast due to out-of-focus signals [4,5]. To overcome these problems, several methods have been proposed for obtaining high-contrast optical sectioning images. For instance, confocal laser scanning microscopy (CLSM) is realized by placing a spatial pinhole filter at the conjugated position of the illumination focus point to reject out-of-focus fluorescence, and the galvo mirror and the motorised stage are often used as a scanning mechanism for acquiring 3D images [6–8]. Line scanning confocal microscopy (LSCM) is established based on a similar principle [9,10], a slit replaces a spatial pinhole in CLSM, and samples are scanned in two directions instead of three to improve the scanning speed. Two-photon excitation microscopy is also an effective way to obtain high-contrast images. With near-infrared two-photon absorption, the scattering in tissues is minimised, and out-of-focus signals are strongly suppressed [11–13]. Light sheet microscopy (LSM) is another powerful optical sectioning technique [14–16]. In LSM, the illumination arm and detection arm are separated orthogonally. Therefore, the imaging objective lens can detect only fluorescent signals from the selective-illuminated plane. This can mitigate phototoxicity and photobleaching and enhance imaging contrast [17–19].

Compared with the above-mentioned methods, structured illumination microscopy (SIM) has a relatively simple configuration. This technique can provide modulated and in-focus images with high contrast and optical sectioning performances [20]. It can be realized by using a digital mirror device (DMD) or a spatial light modulator (SLM) to project periodic patterns on the sample plane [21,22]. To make the optical sectioning easily realized on WFM, Merts' group developed HiLo microscopy using SIM's optical sectioning capability [23–25]. Apart from HiLo's simple configuration, only two images are required for HiLo instead of three for SIM [24]. Using the HiLo principle, we only need one uniform-illuminated image and one structured-illuminated image and use the image processing algorithms to obtain the optical-sectioned image [24]. HiLo is cost-effective by using a coherent laser and a diffuser to create speckles on samples [23–25] or using an incoherent light source and a DMD to project patterns on samples [26,27]. Lim et al. [25] have presented that HiLo's optical sectioning performance is comparable to CLSM [24,25,28]. It can realized 3D image cytometers [29], observe neuron cell activities [30,31], discover 3D cell mechanical properties [32], and enhance retinal imaging quality [33]. HiLo can be easily combined with endoscopy [28,34] and optical scanning microscopy [35,36]. Although HiLo is relatively straightforward compared with other optical sectioning modalities, the cost of its light source, DMD or SLM, and advanced camera sensor is still significant.

Most smartphones have high-performance image sensors with camera lenses, and their costs are quite low. Smartphone-based microscopy is therefore a powerful and low-cost option for bioimaging [37–39], disease detection [40,41], and point-of-care testing applications [42]. Furthermore, smartphones can be easily coupled with different imaging modalities, like bright-field microscopy [43], fluorescence microscopy [44,45], phase microscopy [46–49], and Fourier ptychographic microscopy [50]. However, most of these devices only utilised smartphones' camera lenses and imaging sensors; they still require additional LEDs as external illumination sources, not sufficiently liberating their powerful performances [51–53]. To further simplify smartphone-based microscopy, screen LCD and LED flashlights of smartphones have been used for illumination [50,54].

Here, we propose smartphone-based optical sectioning (SOS) microscopy. To our knowledge, it is the first smartphone-based HiLo microscope that offers low-cost optical-sectioned widefield imaging. The smartphone in our SOS microscope acts as a CMOS sensor to decrease the cost of traditional HiLo microscopes. High-resolution coloured images can be acquired without external colour filters because of the smartphone's Bayer filter and the small pixel size. We made adapters to integrate the smartphone to the main microscope body. A liquid light guide was inserted into the adapter to guide the smartphone's flashlight to the DMD. Meanwhile, a reverse smartphone lens was attached to the camera to conjugate the intermediate image plane into the smartphone's sensor with 1X magnification. The ETL was conjugated to the BPP of the objective lens for realising a telecentric axial scan, which can stabilise SOS's lateral magnification at different depths. The spectrometer tested the smartphone's LED torch, and its broadband spectrum makes it effectively excite fluorescent PS beads with different wavelengths.

SOS has a 571.5 μm telecentric scanning range and an 11.7 μm axial resolution; we successfully used it to acquire high-contrast widefield fluorescent images of 10 μm PS beads with emission wavelengths 465nm, 515nm, and 605nm. Optically sectioned images of accumulated PS beads were also obtained in a 25 μm axial range with a 5 μm step. The proposed system saves more than £7,000 in comparison with a traditional HiLo microscope equipped with a scientific camera sensor and an illumination source. SOS is a low-cost, compact optical sectioning microscope, easy to replicate in a portable style. It has the potential for biomedical research in resource-limited areas.

## 2. Materials and Methods
### 2.1 Optical Setup of the System

Fig. 1(a) presents SOS' optical configuration. A smartphone (iPhone 13Pro, Apple) simultaneously realizes the illumination source and the sensor. For illumination, the LED light from the smartphone is first guided by a liquid light guide (3mm×6', liquid light guide, Edmund Optics). The light from the liquid light guide is filtered by the selected filter and collimated by the Köhler illumination setup (section 2.3 for detail) and illuminates the DMD (DLP LightCrafter 6500, Texas Instruments). The modulated reflective light from the DMD then passes through the dichromatic mirror. L1 is an apochromatic collector lens (MAP10303-A, Thorlabs), and L2 is an aspherical convex lens (ACL3026U-A, Thorlabs). Because each pixel on the DMD is activated along the diagonal line, the DMD is rotated at 45° to maintain the input and reflective light at the same plane (Fig. 1(b)). Beyond that, the illumination light is introduced at 24° from the direction normal to the DMD; the light can be reflected the light at 0°. The DMD plane is conjugated to the sample plane by two 4f systems: L3-RL2 (L3: AC508-180A, RL2: AC508-100A, Thorlabs) and OL-RL1 (OL: UPLanSApo/20X/0.75, Olympus, RL1: AC508-100A, Thorlabs) for structured illumination. The ETL (EL-16-40-TC-VIS-5D-M27, Optotune), with a fast response, acts as a nonmechanical axial scanning device; it is conjugated to the back pupil plane (BPP) of the objective lens. This can guarantee the telecentric setup and maintain the constant lateral magnification during axial scanning (section 2.2 for detail).

The objective lens collects the excited fluorescent signal, and the first image is located at the back focal plane of RL1. Then the image is again imaged at the back focal plane of TL (AC508-180A, Thorlabs) through RL2-TL 4f system. The ETL is located at the confocal plane of RL2, TL, and L3. To acquire the image using a smartphone, we put the same reversed smartphone lens in front of the camera to combine a 1:1 symmetric relay system. So the image can finally be recorded by the smartphone sensor. The smartphone sensor is equipped with the Bayer filter so that coloured images can be directly acquired. Furthermore, we design and fabricate adapters for integrating the smartphone and the liquid light guide to the cage system of the microscope's main body without sacrificing any concentric of the optical path. (Fig. 1(c), section 2.4 for detail)

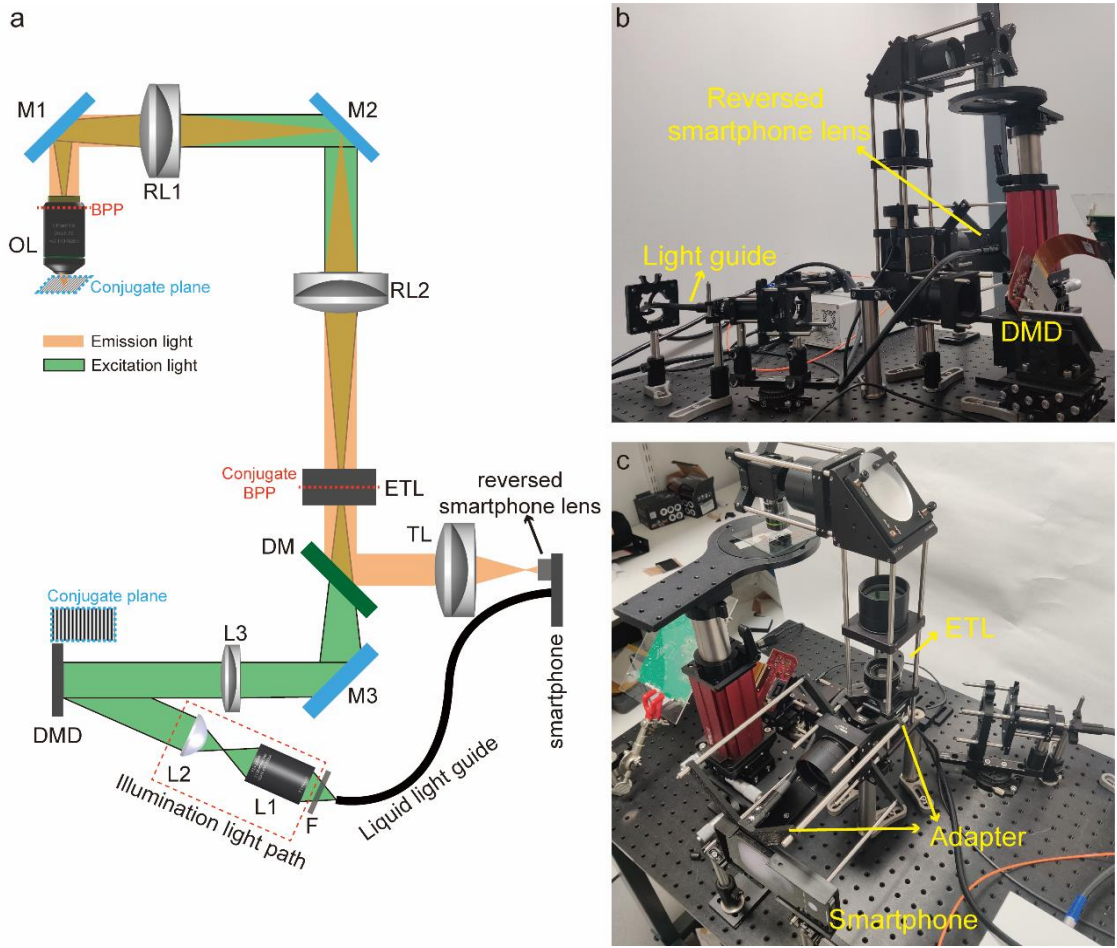

Fig. 1. Optical setup of the proposed SOS microscope. (a) The diagram of the SOS microscope. The DMD plane, which is conjugate to the sample plane, is illuminated by the collimated light. L3, RL2 and RL1, OL consist of two 4f systems. The illumination is introduced at 24° to ensure the DMD can reflect the light at 0°. For telecentric design, the ETL is conjugate to the BPP. The image is recorded by a smartphone equipped with a reversed smartphone lens. (b) and (c) shows the experiment setup. We designed adapters to make the smartphone seamlessly integrate with the SOS main body (section 2.4 for detail). OL: objective lens; M1-M3: mirror; RL1-RL2, L1-L3: lens; F: filter; TL: tube lens; DM: dichromatic mirror; DMD: digital mirror device; ETL: electrical tuneable lens; BPP: back pupil plane.

## 2.2 Telecentric Design and Theoretical Axial Scanning Range

The telecentric design can realized axial scanning with invariant lateral magnification. Here, we used an ETL for axial scanning by tuning the input current to change its surface shape. To maintain lateral magnification at different depths, the entrance and exit pupils should be located infinitely at object and image space, respectively. Therefore, the system setup should be strictly in the 4f configuration.

Fig. 2 illustrates the telecentric design of the imaging path. ETL should be conjugated with the objective lens's back pupil plane (BPP) to obtain the telecentric design. It is hard to put ETL directly at BPP; therefore, we used the 4f configuration (RL1-RL2 lens pair) to relay BPP, at which ETL and BPP are conjugated. Meanwhile, ETL is also located at the front focal plane of the tube lens (TL) to guarantee that the image space is telecentric. It should be noted that the clear aperture size of the conjugate BPP cannot surpass ETL's clear aperture. Here, the BPP diameter of the objective lens is 9 mm and ETL's clear aperture size is 16 mm. We used RL1 and RL2 with both 100 mm focal lengths to combine a 1:1 4f configuration, and its confocal plane coincident with the intermediate image plane I.

To calculate the theoretical scanning range in object space, we should find the relationship between the axial displacement of the object plane and the intermediate image plane. As shown in Fig. 2, $\delta z_+$ and $\Delta z_+$ can be related by:

$$\delta z_+ = \frac{n_o}{n_i M_{OL-RL1}^2} \Delta z_+, \quad (1)$$

$M_{OL-RL1}$ is the lateral magnification of the objective lens and the first relay lens (RL1), $n_o$ and $n_i$ are the refractive indices in the object and the image space, respectively. Here, for simplicity, both of them are unity. The scanning range will be different for the sample's refractive index.

The intermediate image plane I can be tuned by ETL, and we can treat RL2 and ETL as a compound lens. According to Gullstrand's equation [55]:

$$f_{RL2-ETL} = \frac{f_{ETL} f_{RL2}}{f_{ETL} + f_{RL2} - d_{ETL-RL2}}, \quad (2)$$

$f_{RL2-ETL}$ is the focal length of the RL2-ETL-compound lens, $f_{ETL}$ and $f_{RL2}$ are the focal lengthes of ETL and RL, and $d_{ETL-RL2}$ is the distance between ETL and RL2. Since ETL is located at the back focal plane of RL2, Eq. (2) can be rewritten as:

$$f_{RL2-ETL} = \frac{f_{ETL} f_{RL2}}{f_{ETL} + f_{RL2} - f_{RL2}} = f_{RL2}. \quad (3)$$

Therefore, tuning ETL's optical power changes the compound lens's front principal plane instead of changing its focal length. This is equivalent to the axial displacement of the intermediate image plane I. As shown in Fig. 2, $\Delta z_+$ can be written as [55]:

$$\Delta z_+ = \frac{f_{RL2}^2}{f_{ETL}}. \quad (4)$$

Combining Eq. (1) and Eq. (4) we obtain:

$$\delta z_+ = \frac{1}{M_{OL-RL1}^2} \cdot \frac{\phi_{ETL}}{\phi_{RL2}^2}, \quad (5)$$

$\phi$ denote the corresponding optical power.

Therefore, the scanning range in the object space can be written as:

$$\delta z = |\delta z_-| + |\delta z_+| = \frac{1}{M_{OL-RL1}^2} \left( \left| \frac{\phi_{ETL-max}}{\phi_{RL2}^2} \right| + \left| \frac{\phi_{ETL-min}}{\phi_{RL2}^2} \right| \right), \quad (6)$$

$\phi_{ETL-min}$ and $\phi_{ETL-max}$ are the minimum and maximum optical power of ETL.

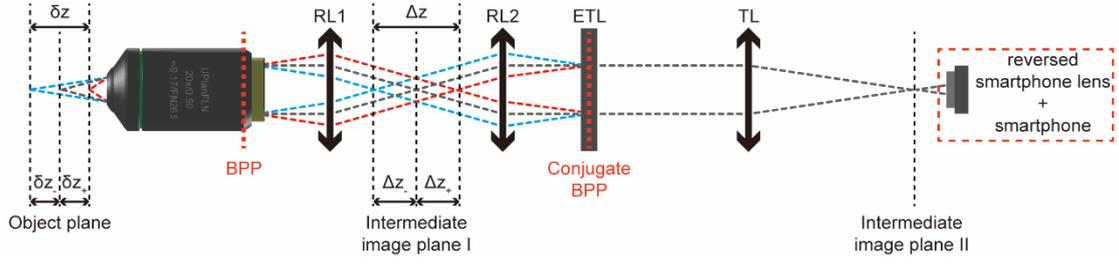

Fig. 2. Telecentric imaging optical path. We configure RL1 and RL2 as a 4f system to relay the objective lens' back pupil plane to ETL. The red and blue lines depict the minimum and maximum axial scanning ranges.

## *2.3 Illumination Path Design*

Fig. 3 shows the exact illumination path for the HiLo microscopy. The light guide's light must be collimated, and the final beam diameter *D* should fill the DMD's active area. Furthermore, this diameter *D* cannot be too large to prevent illumination energy loss. Therefore, we set *D* to be the diagonal length of DMD's active area. We applied the Köhler setup to acquire high-quality illumination. We used L1 as a collector to image the light source to the front focal plane of the condenser (L2), and the final collimated light illuminated the DMD. To choose appropriate elements for L1 and L2, we can find the relationship between the illumination half angle *u* and beam diameter *D* according to the Gaussian optics. The magnification of the collector (L1) can be calculated as:

$$m_c = \frac{n_1 \sin u}{n_1' \sin u'}, \qquad (7)$$

where $n_1$ and $n_1'$ are the refractive index in the object and image space of L1, respectively.

When the light is collimated by L2, the relationship between $u'$ and *D* is:

$$\sin u' = \sin[\arctan(\frac{D}{2 f_{condenser}})], \qquad (8)$$

$f_{condenser}$ is L2's focal length. The medium is air. Combining Eq. (7) and Eq. (8), we can obtain:

$$\sin u = m_c \sin[\arctan(\frac{D}{2 f_{condenser}})], \qquad (9)$$

where the angle $u$ can be obtained from experiments (Part 3.2 for detail) and *D* should be slightly longer than the diagonal length of the DMD active area, which is 12mm.

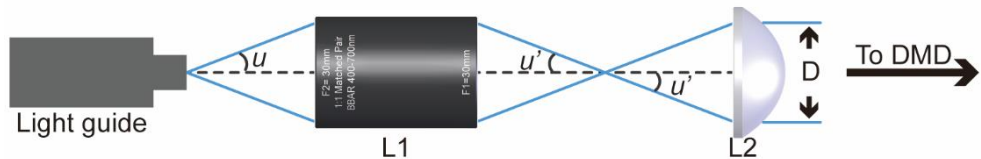

Fig. 3. Illumination light path. After calculating and experiments, we choose appropriate L1 (MAP10303-A, Thorlabs) and L2 (ACL3026U-A, Thorlabs). L1 has 1X magnification, and the NA of L2 is 0.55.

## 2.4 Smartphone Adapter Design

We employed CAD design (Autodesk, Inventor Professional 2020) and 3D-printed two adaptors. Fig. 4(a) shows the connections between these adaptors. The two holders was designed to accommodate the iPhone. These adapters are designed similarly to the Thorlabs cage system adapter. So that the smartphone can be easily combined with the microscope's main body without losing any concentricity. Fig. 4(b) illustrates the smartphone holder with the light guide and the reversed smartphone lens.

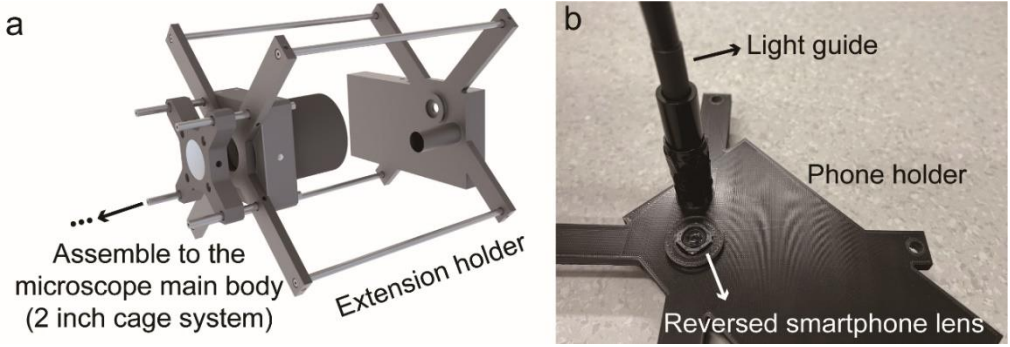

Fig. 4. Home-made adaptors. (a) The diagram of harnessing the smartphone to the microscope's main body. (b) The light guide and the reversed smartphone lens are aligned through the smartphone holder.

## 2.5 HiLo Algorithms

The detailed theoretical principle of HiLo has been derived by Mertz et al. [23–25,56,57]. Briefly, in HiLo, we use two images captured under uniform and structured illumination to extract the in-focus high-frequency components (IHC) and in-focus low-frequency components (ILC) and generate an optical-sectioning image $I_{HiLo}(x, y)$:

$$I_{HiLo}(x, y) = \eta I_{Lo}(x, y) + I_{Hi}(x, y), \quad (10)$$

where $I_{Lo}(x, y)$ is the in-focus low-frequency image and $I_{Hi}(x, y)$ is the in-focus high-frequency image, and $x$ and $y$ are spatial coordinates at the image plane. The parameter $\eta$ is used to avoid the discontinuity at all frequencies, calculated by [34]:

$$\eta = \frac{HP_{Kc}}{LP_{Kc}}, \quad (11)$$

where $HP_{Kc}$ and $LP_{Kc}$ are Gaussian high-pass and low-pass filters, respectively, and $K_c$, the cut-off frequency, should be less than or equal to the frequency of the structured illumination pattern.

According to the wide-field microscope's optical transfer function (OTF), high-frequency features decay rapidly, defocused from the image plane [58]. Thus, we can apply a high-pass filter on the uniformly-illuminated image to obtain ICH:

$$I_{Hi}(x, y) = \mathfrak{I}^{-1}\{HP_{Kc}\{\mathfrak{I}[I_u(x, y)]\}\}, \quad (12)$$

where $\Im$ and $\Im^{-1}$ are two-dimensional Fourier transform and inverse Fourier transform, respectively. $HP_{Kc}$ denotes a Gaussian high-pass filter with the cutoff frequency $Kc$. $I_u$ is the image under uniform illumination, and it can be divided into the in-focus term $I_{focus}$, and the defocused term $I_{defocus}$:

$$I_u(x, y) = I_{focus}(x, y) + I_{defocus}(x, y) \qquad (13)$$

Similarly, the structurally-illuminated image under sinusoidal illumination is:

$$I_s(x, y) = I_{focus}(x, y) + M \sin(2\pi kx) I_{focus}(x, y) + I_{defocus}(x, y), \qquad (14)$$

where $M$ is the modulation depth and $k$ is the pattern spatial frequency. The sinusoidal pattern can modulate in-focus images, whereas defocused images does not.

To extract the ILC from $I_u(x, y)$, we should use a weighting function to reject defocused low-frequency components. Since the image modulated by the sinusoidal pattern have the maximum contrast at the focal plane, we can relate this weighting function with the contrast. This weighting function can be obtained by using the rectified subtraction method [28], the single-sideband demodulation method [34], and the local contrast method [24]. To remove the sample-induced contrast and obtain a better axial resolution, we use the difference image with a band-pass filter for the contrast [31]:

$$C(x, y) = std \left\langle \Im^{-1}\{BPF\{\Im[|I_u(x, y) - I_s(x, y)|]\}\} \right\rangle, \qquad (15)$$

where $std$ represents the stand deviation operation and $BPF$ is a 2D Gaussian band-pass filter. The band-pass filter can remove the DC spectral components, which is not defocus. We can enhance the optical sectioning performance of low-frequency components by tuning the bandpass filter's width, and it can be denoted as:

$$BPF(k_x, k_y) = \exp(-\frac{k_x^2 + k_y^2}{2\sigma^2}) - \exp(-\frac{k_x^2 + k_y^2}{\sigma^2}), \qquad (16)$$

where $(k_x, k_y)$ is the spectral coordinate and $\sigma$ is the band-pass filter's width.

We can obtain weighted uniformly-illuminated images without $I_{defocus}$ by multiplying the uniformly-illuminated image with $C(x, y)$. Then, we apply a low-pass filter to the weighted

uniformly-illuminated image to diminish structurally-illuminated-induced sinusoidal noise to acquire ILC:

$$I_{Lo}(x, y) = \mathfrak{I}^{-1}\{LP_{Kc}\{\mathfrak{I}[C(x, y)I_u(x, y)]\}\}, \qquad (17)$$

where $LP_{Kc}$ is the low-pass filter with cutoff frequency $Kc$.

## 3. Results
### 3.1 Performance of Axial Scanning

The theoretical axial scanning range can be obtained from Eq. (6). The effective ETL current ($I$) ranges from -250 mA to 250 mA, corresponding to the optical power from -3.3 to 3.5 dpt. The lateral magnification of the lens pair OL-RL1 is 11.1, and the optical power of RL2 is 10 dpt. The theoretical axial scanning range is calculated as 561 μm.

The experimental axial scanning range is also measured to compare with the theoretical results. We uploaded a uniform pattern onto the DMD to generate wide-field illumination. A resolution target (R3L3S5P, Thorlabs) was imaged at the different axial positions by tuning the current of the ETL and the z-axis translation stage. In the experiment, we adjusted $I$ to be -250, 125, 0, 125, and 250 mA. After each current adjustment, we axially moved the target by adjusting the translation stage (Vertical Translation Stages, Edmund Optics) to get a clear image, and the relationship between $I$ and the axial displacement ($d$) can be found (Fig. 5(c)). The translation stage here is for finding relationship between the axial movement and the ETL current. It can be exempted during experiments.

Fig. 5(a) shows five in-focus images at different depths; the central image is the in-focus image without tuning the $I$ (0mA). All images were captured without any horizontal or vertical movements. To testify the telecentric property of the proposed system during axial scanning, as shown in Fig. 5(a), we traced three line profiles at the same position when $I$ were -250mA, 0mA, and 250mA. As shown in Fig. 5(b), the size of each circle fixed, meaning that the lateral magnification is constant. Fig. 5(c) shows the ETL current versus the axial displacement. We set the axial position when $I$ is zero as a reference plane ($d$ is zero). The measured axial scanning range is 571.5 μm, close to the theoretical value of 561 μm with a 2% error.

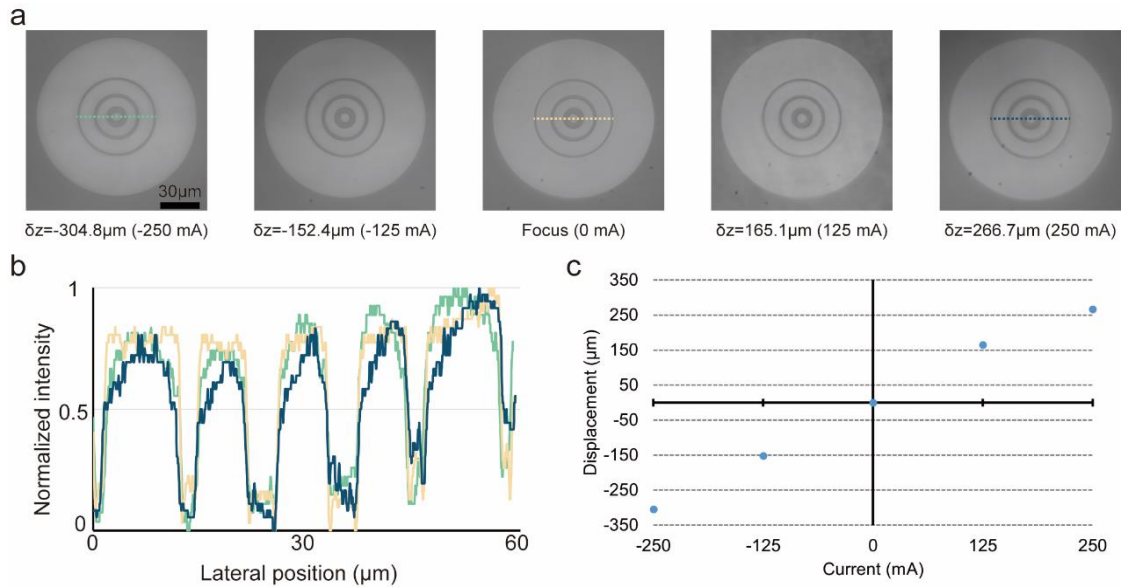

Fig. 5. Performance of axial scanning and telecentric property. (a) In-focus images at different depths. The ETL current is first changed, then the target is moved axially to acquire a clear picture. (b) Corresponding line profiles in Fig. 5(a). (c) Relationship between the ETL current and axial displacement.

## 3.2 Characterization of the Illumination Path
### a. Radial Intensity Distribution
According to Eq. (9), the light source's radial intensity distribution decides how to choose the appropriate optical components to illuminate the whole DMD area with proper beam diameter. We contacted the low-loss liquid light guide (5mm * 6' UV, Edmund Optics) to the smartphone's LED and tested the output illumination radial intensity distribution, as shown in Fig. 6(c). We also compared the radial intensity distribution directly irradiated from the smartphone shown in Fig. 6(a). Figures 6(b) and (d) show the corresponding distributions and corresponding distributions in the polar coordinate system. The light guide can easily guide the illumination source to the aimed position. Besides, the power distribution from the light guide is more focused (14.2° FWHM).

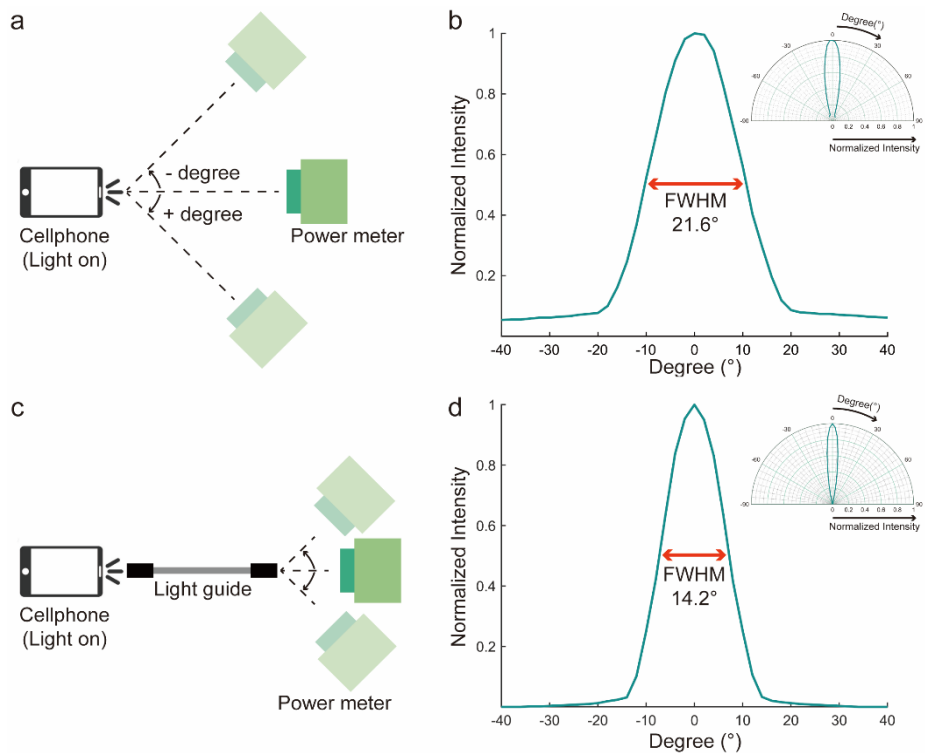

Fig. 6. The measured radial intensity distribution by the power meter (LASERPOINT, Italy). Intensity distribution tests with (a) and without (c) the light guide. (b) The distribution of (a). (d) The distribution of (c). FWHM means the full-width-half-maximum. The subplots in (b) and (d) are in the polar coordinate system.

### b. Spectral Intensity Distribution
We used the spectrometer (HR2000, Ocean Optics) to test the smartphone LED's spectrum. Fig. 7 shows the spectrum range of the iPhone 13pro LED. Like most white LEDs, the peak emission appears at ~450nm wavelength, and the normalised intensity in the visible spectrum range is higher than 50% except the dip around 480 nm.

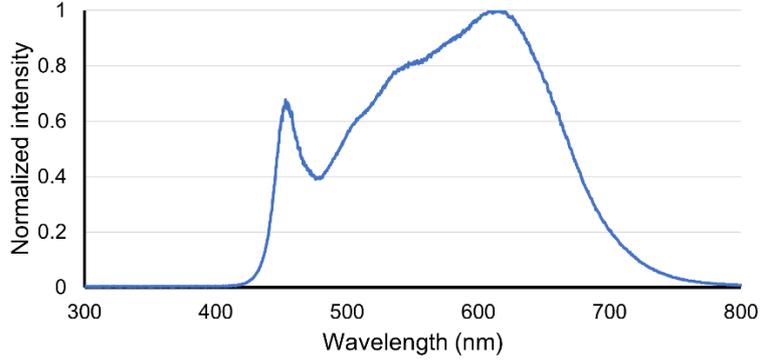

Fig. 7. The spectrum of the iPhone 13pro LED.

*3.3 Optical Sectioning Capability*

To test SOS' optical sectioning capability, we put a silver-coated mirror (PFR10-P01, Thorlabs) on the sample stage and projected DMD patterns with different periods to its surface. Since the pattern contrast decreases with defocus, we extracted contrast maps from pattern images and quantify the optical sectioning capability accordingly [28]. We used green light (532nm) as reference [59]. An emission filter (FL532-3, Thoralbs) was inserted in front of the liquid guide. The focal plane was axially scanned with a 5μm step for a 15.12μm/lp pattern and a 10μm step for 75.6μm/lp and 241.92μm/lp patterns. The normalised contrast for each axial image is calculated by $C = (I_{max} - I_{min})/(I_{max} + I_{min})$. Fig. 8(a) shows the relationship between the normalised contrast and the axial position for three patterns with different frequencies. A higher frequency shows better optical section capability. This is because the higher spatial frequency modulation of the OTF in widefield microscopy decays with defocus more quickly, enabling better optical sectioning power. In this study, the DMD's pixel size is 7.56μm, so the highest frequency is 15.12μm/lp. Fig. 8(b) shows images with three frequencies captured at three axial positions. The pattern structure at 15.12μm/lp almost vanishes when the defocus distance is 10μm, whereas the periodic structure at 241.92μm/lp is still observable at the defocus distance 100μm.

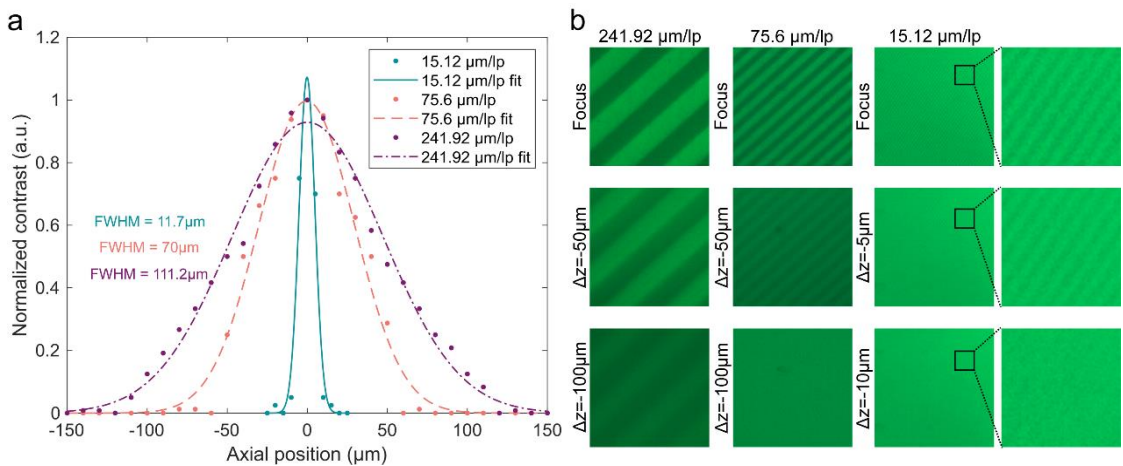

Fig. 8. SOS' optical sectioning capability. (a) The axial contrast profiles of DMD patterns for three spatial frequencies were obtained by imaging the reflected signal from the mirror sample. We used the Gaussian function for each experimental dataset. The Gaussian function's FWHM is also shown. (b) The images with three frequencies in (a) are captured at focus, and two defocus planes.

### 3.4 Fluorescent beads imaging

We used three fluorescent beads with 465 nm, 515 nm, and 605 nm excitation wavelengths (FluoSpheres, Thermo Fisher) to test SOS' imaging performances. The bead's size is 10μm in diameter. The samples were diluted ten times with denoised water and set on microscopic glass slides. For structured illumination, the DMD pattern period was set to 6 pix/lp (45.36μm/lp). For uniform illumination, all pixels on DMD were turned 'on'. Figure 9 shows the captured images of fluorescent bead samples. The widefield images were imaged under uniform illumination, and the SI images were under structured illumination. We used the HiLo algorithm to obtain the final HiLo images by taking both widefield and SI images. To excite different fluorescent beads, we insert different optical filters into the illumination path which was shown in Fig. 1(a), letter F (Fig. 9(a), FBH430-10; Fig. 9(b), FBH500-40; Fig. 9(c), FBH580-10, Thorlabs). Because of the Bayer filter on the smartphone sensor, emission filters are unnecessary. Fig. 9(d)- (f) shows corresponding line profiles in Fig. 9(a)- (c).

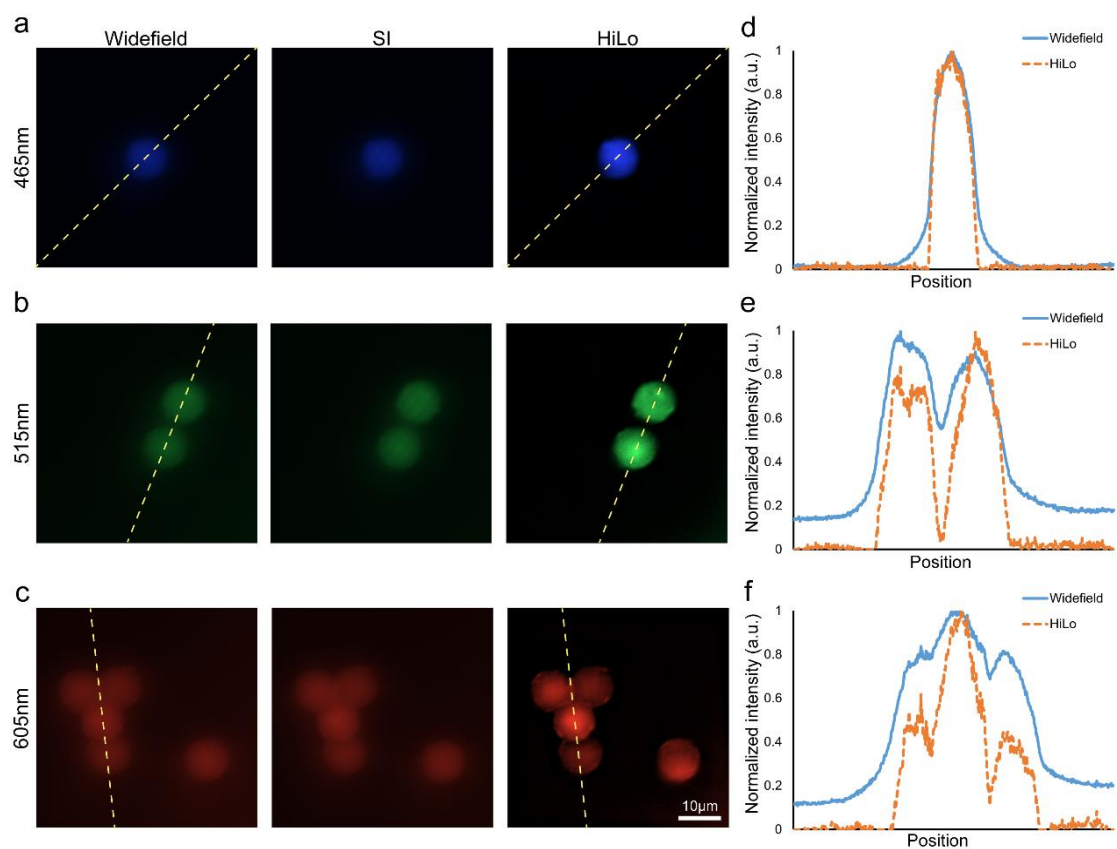

Fig. 9. Fluorescent bead images at different wavelengths. (a) The bead samples with 430nm excitation and 465nm emission. (b) The bead samples with 505nm excitation and 515nm emission. (c) The bead samples with 580nm excitation and 605nm emission. Corresponding line profiles of widefield and HiLo images are shown in (d)-(f). SI: Structured illumination.

### 3.5 Optically sectioned imaging of accumulated fluorescent beads

To test SOS' optical sectioning performances, the fluorescent beads (FluoSpheres, Thermo Fisher, 505/515) of different layers were optically sectioned and then imaged. We dipped the 10μL bead solution without dilution onto the microscopic glass side and waited until it was dry. Therefore, beads start to accumulate and flow much slower. $I$ was set to -10.5, -7.0, -3.5, 0.0, 3.5, 7.0 mA , and the focus plane from -15 μm to -10 μm with a 5 μm step. At each step, beads were illuminated by uniform and structured illumination (45.36 μm/lp). Fig. 10 shows widefield, structured illumination, and HiLo-processed images of accumulated beads at different depths. The HiLo images show much better optical sectioning capacity compared with widefield images. Out-of-focus signals are significantly suppressed.

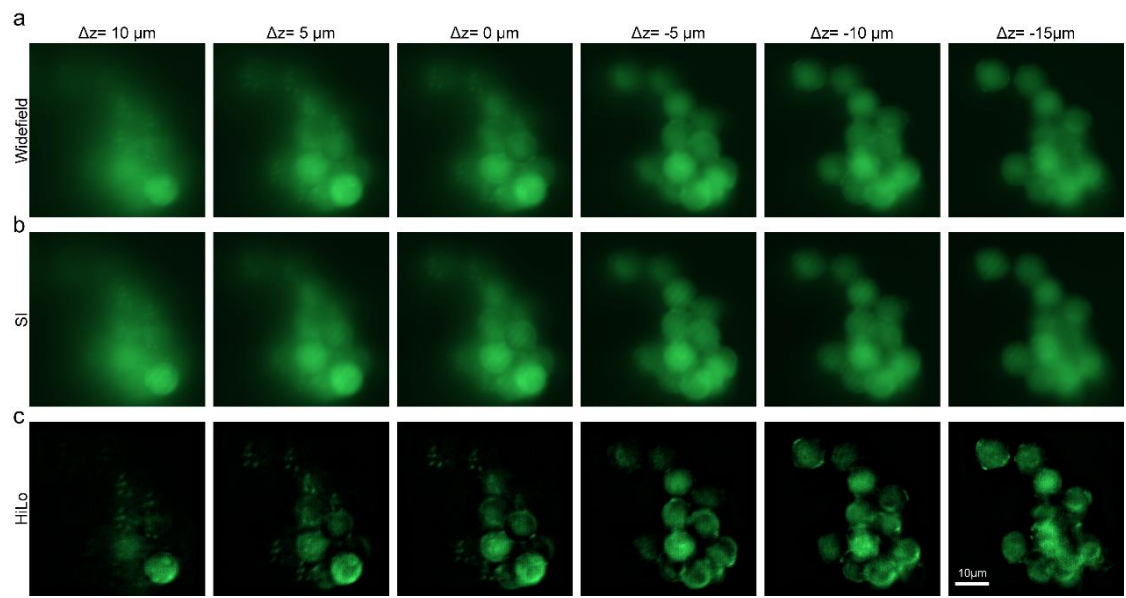

Fig. 10. Images of accumulated fluorescent beads at different depths. (a) widefield images. (b) SI images. (c) HiLo processed images.

## 4 Discussions and Future Aspects

In resource-limited areas, using microscopes for biomedical research and disease diagnosis is difficult due to insufficient resources. Especially for advanced optical sectioning microscopes, light sources, scientific sensors, precise translation stages and optical components are usually costly. In this article, we proposed a smartphone-based HiLo optical sectioning microscope, SOS, to offer a low-cost solution with a smartphone and an electrically tunable lens. Table. 1 summarizes the components used in our SOS or traditional HiLo and the corresponding costs. If the other components (objective lens, mirror, cage rod, lenses, filter) in SOS and traditional HiLo systems are all the same, SOS can save around £7,435.

SOS's main structure was precisely designed and built by the compact cage system. We designed and fabricated adapters for combining the smartphone with the cage structure. We used a liquid light guide to guide the smartphone torch's light to the DMD for structured illumination. The advantages: The liquid light guide can freely direct light to aimed positions with a bit loss and more focused illumination, and the light source image disappears during the transmission inside the light guide.

The ETL in our SOS is an axial scanner for selecting sample planes and adjusting the focal plane. The ETL is conjugated to the BPP of the objective lens by a 4f configuration. This telecentric setup can maintain the lateral magnification regardless of the axial plane's position. Therefore, the field of view and the resolution are maintained. Moreover, Z-stack images can be directly generated without any post-processing. The broadband LED light source can effectively excite different PS beads.

The proof-of-concept SOS protoype can successfully realize section and image 10 μm PS fluorescent beads, meaning that it has great potential for imaging more sophisticated fluorescently labelled biological samples. In the future, we will use SOS to conduct more biological experiments with thicker biological samples (such as fluorescently labelled zebrafish) to develop a compact portable SOS. We will design a whole mechanical structure and assemble all components, and the cage system can be exempted. We will also develop a standard adapter that combines different types of smartphones.

Table .1 Costs of the proposed SOS and traditional HiLo systems

|  | **SOS** | **Traditional HiLo** |
|---|---|---|
| **Light source** | N/A | Tungsten-Halogen Source: ~ £5,000 |
| **Camera sensor** | N/A | 12 MP colour CMOS camera: ~ £2,000 |
| **Smartphone** | Iphone 13pro: ~ £615 | N/A |
| **Axial translation stage** | Electrically tunable lens: ~ £950 | Motorised translation stage: ~ £2000 |
|  | Total: ~ £1,565 | Total: ~ £9,000 |

## 5  Conclusion

We proposed a smartphone-based optical sectioning (SOS) microscope based on the HiLo principle. Compared with traditional HiLo, it can save more than £7,000. In SOS, the illumination source and camera sensor can be realized by a smartphone. We use an ETL for the axial translation stage, realising axial scanning and focus plane selection. The telecentric setup can maintain the lateral magnification regardless of the axial plane's position, and post-processing for imaging registration is exempted. SOS offers a 571.5 μm axial range, close to the theoretical value (561 μm) with a 2% error. The axial resolution can reach 11.7 μm. We successfully used SOS to image different PS beads, and the optically sectioned images of accumulated beads were also recorded in the 25 μm axial range with a 5 μm step.

**Declaration of competing interest**
There are no conflicts to declare

**CRediT authorship contribution statement**
**Ziao Jiao**: Investigation, conceptualization, validation, hardware design, sample preparation, data analysis, article writing and editing. **Mingliang pan**: Hardware design, data analysis. **Khadija Yousaf** and **Daniel Doveiko**: Sample preparation. **Michelle Maclean** and **David Griffin**: Spectrum testing. **Yu Chen**: Sample preparation and supervision. **David Day Uei Li**: Project supervision & administration, concept presentation, article writing and editing, & funding acquisition.

**Data availability**
Data underlying the results presented in this paper are not publicly available but may be obtained from the authors upon request.

**Acknowledgement**

The research has been supported by the Engineering and Physical Sciences Research Council under EPSRC Grant: EP/L01596X/1, the Royal Society of Edinburgh, and the China Scholarship Council. Daniel Doveiko would like to thank PQ Corporation and EPSRC for the PhD studentship (2629179).